\documentclass[manuscript]{aastex}

\shorttitle{Twist and Dip Shear}
\shortauthors{Gosain et al.}

\begin{document}

\title{Evolution of twist-shear and dip-shear during  X-class flare of 13 December 2006: Hinode observations}
\author{Sanjay Gosain and P. Venkatakrishnan}
\affil{Udaipur Solar Observatory,Physical Research Laboratory,
Udaipur, Rajasthan- 313 001,
INDIA}

\begin{abstract}
The non-potentiality (NP) of the solar magnetic fields is
measured traditionally in terms of magnetic shear angle i.e.,
the angle between observed and potential field azimuth. Here,
we introduce another  measure of shear  that has not been
studied earlier in solar active regions, i.e. the one that is
associated with the inclination angle of the magnetic field.
This form of  shear, which we call as the  ``dip-shear", can be
calculated by taking the difference between the observed and
potential field inclination. In this {\it Letter},  we study
the evolution of dip-shear as well as the conventional
twist-shear in a $\delta$-sunspot using high-resolution vector
magnetograms from {\it Hinode} space mission.  We monitor these
shears in a penumbral region located close to flare site during
12 and 13 December 2006. It is found that: (i) the penumbral
area close to the  flaring site  shows high value of
twist-shear and dip-shear as compared to other parts of
penumbra, (ii) after the flare the value of  dip-shear drops in
this region while the  twist-shear in this region tends to
increase after the flare, (iii)  the dip-shear and twist-shear
are correlated such that pixels with large twist-shear also
tend to exhibit large dip-shear, and (iv) the correlation
between the twist-shear and dip-shear is tighter after the
flare. The present study suggests that monitoring twist-shear
during the flare alone is not sufficient but we need to monitor
it together with dip-shear.

\end{abstract}

\section{Introduction}
 The deviation of an active region (AR) magnetic field from its potential
configuration can arise  due to photospheric footpoint motions
and/or during flux emergence. However, studies with the
high-quality data from space and ground based observatories
over the last decade suggest that the latter is more dominant
mechanism
\citep{Wheatland2000,Pascal2002,Falconer2002,Leka2003,Schrijver2005,Jing2006,Schrijver2007}.
The deviation of magnetic field from potential configuration is
called non-potentiality (NP) of the field. The magnetic energy
of an active region (AR) in excess of the potential magnetic
energy is called free magnetic energy \citep{Low82}. The free
magnetic energy of an AR, or a portion of it, is released when
flare and/or Coronal Mass Ejections (CMEs) is triggered due to
the instability or non-equilibrium \citep{Priest2002}. The
energy released in the flare is then limited by the amount of
free energy or magnetic NP of the AR. Therefore, it is
important to characterize the NP of the AR magnetic field in
order to predict  the intensity of the flare and/or CME.

 Traditionally, the  so-called magnetic shear i.e.,  the angle
between the observed ($\psi_{o}$) and the potential
($\psi_{p}$) field azimuths, is used to characterize the NP of
the active region magnetic field and is measured as
$${\rm  \Delta\psi=acos ~\frac{ \vec{B_t^{o}} \cdot \vec{B_t^{p}}}{|B_t^{o}||B_t^{p}|}}$$
where ${\rm \vec{B_t^{o}}}$ and ${\rm \vec{B_t^{p}}}$ are the
observed and potential transverse field vectors.  The term
``magnetic shear", used here and in similar studies on
solar-magnetism, really refers to ``observational shear", which
is not to be confused with the actual ``shear"--a microscopic
property in fluid dynamics. In what follows, we will use the
term magnetic shear for traditional reasons and by this term we
would always mean the observational shear. Various forms of
this parameter are used namely, mean shear \citep{Mona1984},
weighted mean shear \citep{wang1992}, spatially averaged signed
shear \citep{sanjiv2009}, most probable shear \citep{Moon2003}
etc.  It is well known that the polarity inversion line (PIL)
in active regions bearing highly sheared magnetic fields
 are the potential sites for flares \citep{Mona1984,Hagyard1990}.
 These PILs are generally characterized by dark filaments in the images taken in chromospheric hydrogen alpha
line \citep{Zirin1973}.

 It may be noted that all of the shear parameters mentioned
above measure the twist-shear component of  the shear which is
measured in the horizontal plane. However, one can also measure
the shear of a magnetic field in the vertical plane. This type
of shear can be  called as the  ``dip-shear" (we choose the
term ``dip" because it is synonymous to the dip angle measured
in geomagnetism), and can be measured by taking the difference
between the observed ($\gamma_{o}$) and the potential field
inclination angle ($\gamma_{p}$) i.e.,
$\Delta\gamma=(\gamma_{o}-\gamma_{p})$.

Physically, the dip-shear can be understood in terms of
azimuthal currents, in the same way as the twist-shear is
understood in terms of axial currents. In our knowledge, the
dip-shear  has not been studied earlier in solar active
regions. Henceforth, we will call the parameter $\Delta\psi$
and $\Delta \gamma$ and as twist-shear and dip-shear,
respectively. The larger the value of these angles the larger
will be the  NP of the AR. It may be noticed that, unlike
twist-shear, the dip-shear  is not affected by the 180 degree
azimuth ambiguity, provided that the active region is observed
close to the disk center. This is because the dip-shear depends
upon the inclination angle of the magnetic field which can be
measured without ambiguity \citep{Metcalf2006}.

In this {\it letter}, we  study the evolution of twist-shear
and dip-shear in a penumbral region located close to the
flaring site in AR NOAA 10930.   We use sequence of
high-quality vector magnetograms observed by the {\it Hinode}
space mission. This active region was in a $\delta$-sunspot
configuration  which led to a X3.4 flare and a large CME during
02:20 UT on 13 December 2006. The flare was quite powerful and
the white light flare ribbons along with impulsive lateral
motion of the penumbral filaments were observed
\citep{Gosain09}.

 Evolution of the twist-shear and dip-shear together shows
interesting patterns which can be distinguished in the
pre-flare and post-flare stages. In general we find that (a)
the regions with high twist-shear also exhibit high dip-shear,
(b) the penumbral region close to the flare site shows high
twist-shear and dip-shear, and (c) twist-shear and dip-shear
studied together can be used to study flare related changes in
the active regions.


The paper is organized as follows. The observational data and
the  methods of analysis are described in section 2. The
results are presented in section 3 and the discussions and
conclusions are made in section 4.


\section{Observational Data and Methods of Analysis}
\subsection{Hinode observations}
A sunspot with $\delta$ configuration was observed in AR NOAA
10930 during 12-13 December 2006 by the Spectro-Polarimeter
(SP) instrument \citep{Lites2007,Ichimoto2008} with Solar
Optical Telescope (SOT) \citep{Tsuneta2008} onboard {\it
Hinode} satellite \citep{Kosugi2007}. The SP obtains the Stokes
profiles, simultaneously in Fe I 6301.5 \AA~ and 6302.5 \AA~
line pair. The spectro-polarimetric maps of the active region
are made by scanning the slit across the field-of-view. This
takes about an hour to complete one scan. We choose  a sequence
of six SP scans from 12 December 2006 03:50 UT to 13 December
2006 16:21 UT  when the sunspot was located close to the disk
center with heliocentric distance ($\mu$) of  0.99 and 0.97,
respectively. The scans were taken in ``Fast Map" observing
mode with following characteristics: (i) Field-of-view (FOV)
295 x 162 arc-sec, (ii) integration time of 1.8 seconds and
(iii) pixel-width across and along the slit of 0.32 and 0.29
arc-sec, respectively. The Stokes profiles were then fitted to
an analytic solution of Unno-Rachkovsky equations
\citep{Unno1956,Rachkovsky1973} under the assumptions of Local
Thermodynamic Equilibrium (LTE) and Milne-Eddington model
atmosphere \citep{Landi1982} with a non-linear least square
fitting code called {\it HeliX} \citep{Lagg04}. The physical
parameters of the model atmosphere retrieved after inversion
are  the magnetic field strength, its inclination and azimuth,
the line-of-sight velocity, the Doppler width, the damping
constant, the ratio of the line center to the continuum
opacity, the slope of the source function and the source
function at $\tau$ = 0. We fit a single component model
atmosphere along with a stray light component. The inversion
code {\it HeliX} is based upon a reliable genetic algorithm
\citep{Charbonneau1995}. This algorithm, although slower, is
more robust than the classical Levenberg-Marquardt algorithm in
the sense that the global minimum of the merit function is
reached with higher reliability \citep{Lagg04}.

The vector magnetograms obtained after inversion, were first
solved for 180 degree azimuth ambiguity by using the acute
angle method \citep{Harvey69} and then transformed from the
observed frame (image plane) to local solar frame (heliographic
plane) using the procedure described in \cite{pvk1988}. The
potential field was computed from the line-of-sight field
component by using the Fourier transform method
\citep{Alissandrakis1981,Gary1989}. The IDL routine used for
potential field computation is \verb+fff.pro+ which is
available in the NLFFF (Non-linear force free field)
 package of the SolarSoft library.  The continuum
intensity images of the sunspot, corresponding to the sequence
of scans used,  are shown in figure 1, with the transverse
field vectors overlaid on it. The two magnetograms were aligned
using the cross-correlation technique applied to the continuum
image of the sunspot. A black rectangle is overlaid on these
images to show the location of the region where we study the
evolution of twist-shear and dip-shear. The location of this
black rectangle is chosen with the help of a co-aligned G-band
filtergram observed from Hinode Filtergraph (FG) instrument
during flare. This G-band image is shown in the left panel of
figure 2. The flare ribbon is marked by `+' symbols and the
black rectangle of figure 1 is shown here also. The flare
ribbons sweep across the rectangular box during 02:20 to 02:26
UT. This indicates that the rectangle is chosen such that it
samples the penumbra which is very close to the flaring site.
The right panel of figure 2 shows the longitudinal magnetogram
in order to indicate the location of rectangle (shown here with
white color) with respect to the PIL. The maps of dip-shear
$\Delta \gamma$ and twist-shear $\Delta \psi$ for the sequence
of vector magnetograms are shown in figure 3 and 4
respectively. In figure 5 we show the distribution of dip-shear
$\Delta\gamma$ and twist-shear $\Delta\psi$ inside the black
rectangle and its evolution with time.

\section{Results}
\subsection{Distribution and Evolution of Non-Potentiality in NOAA 10930}
\subsubsection{Dip-Shear}
The maps of  dip-shear  $\Delta \gamma$  for the entire
sequence of vector magnetograms covering the pre-flare (panels
(a)--(c)) and post-flare (panels (d)--(f)) phases are shown in
figure 3. The value of field inclination $\gamma$ is measured
with respect to local solar vertical direction and ranges from
0 to 180 degrees. For purely vertical positive (negative)
polarity field the value of $\gamma$ corresponds to  0 degrees
(180 degrees). The black rectangle in figure 3 corresponds to
negative polarity field. Therefore, the positive value of
dip-shear $\Delta \gamma$ inside this rectangle means that the
observed field is more vertical than potential field. The
magnitude of dip-shear $\Delta \gamma$ can be judged with the
aid of colorbar at the bottom of figure 3.

It may be noticed that: \\
(i) the value of dip-shear $\Delta \gamma$ is large inside the
rectangle as compared to other penumbral locations. \\
(ii) in the pre-flare (panels (a)--(c)) phase the dip-shear
$\Delta \gamma$ consistently  has a large magnitude which
decreases in the post-flare  phase (panels (d)--(f)).

\subsubsection{Twist-Shear}
 The field azimuth has been solved for 180 degree azimuth
ambiguity by using the acute angle method \citep{Harvey69} and
the projection effects have been removed by the application of
vector transformation \citep{pvk1988}. This azimuth ambiguity
resolution method works well in the regions where the angle
$\Delta\psi$ is less or greater than 90$^\circ$. For regions
where $\Delta\psi$ reaches value close to 90$^\circ$ such as
along  parts of polarity inversion line (PIL) in flaring active
regions, the accuracy of the method is poor. This is why we
choose a rectangular box for studying the evolution of
twist-shear and dip-shear sample the penumbra close to flaring
site and at the same time stay away from the PIL where the
acute angle method may have problems in resolving the azimuth
ambiguity.

 The maps of twist-shear
$\Delta \psi$  for the entire sequence of vector magnetograms
covering the pre-flare (panels (a)--(c)) and post-flare (panels
(d)--(f)) phases are shown in figure 4. The value of field
azimuth $\psi$ is measured with respect to the positive X-axis
 and is positive in the anti-clockwise direction. The
magnitude of twist-shear $\Delta \psi$ can be judged with the
aid of colorbar at the bottom of figure 4. However, for the
present study the sign of shear angle is not important, so we
shall focus on its magnitude. It may be noticed that the value
of twist-shear $\Delta \psi$ is large inside the rectangle and
adjacent PIL as compared to other penumbral locations. However,
the flare related changes are not so discernable to eye as
compared to dip-shear in figure 3.

\subsubsection{Evolution of Dip-Shear and Twist-Shear}
 In figure 5 we show the scatter between the  dip-shear
$\Delta\gamma$ and  twist-shear $\Delta\psi$ for the pixels
within the black rectangle shown in previous figures. The
panels (a) to (c) correspond to pre-flare while the panels
i.e., (d) to (f) correspond to post-flare phase.

It may be noticed that : \\
(i) The distribution of dip-shear $\Delta\gamma$ and twist-shear $\Delta\psi$ in panels (a)-(c) is different from the distribution in panels (d)-(f).\\
(ii) The dip-shear increases before the flare (panels (a)-(c))
but twist-shear tends to decrease at the same time. \\
(iii) The dip-shear and twist-shear are in general correlated, i.e. the pixels with large dip-shear also have large twist-shear.\\
(iv) The most important change can be noticed after the flare,
i.e., between panels (c) and (d). After the flare, (panel (d))
the dip-shear decreases significantly while twist-shear
increases. However, now both shear components show less
dispersion i.e., follow a tight correlation.\\
(v) In panels (d)-(f) the two shears maintain smaller
dispersion but dip-shear starts to increase once again.  This
increase suggests that the non-potentiality was building up
again in the active region. It may be noted that the flaring
activity continued in this region on the next day i.e., 14
December 2006 also, with another X-class flare occurring at
about 22:00 UT.

\section{Discussion and Conclusion}
%
 \cite{Hugh2000} conjectured that the free-energy is stored
in non-potential magnetic loops that are stretched upwards and
the free-energy release during the flare must be accompanied by
sudden shrinkage or implosion in the field. Also, it is
predicted that after the flare the field should become more
horizontal \citep{Hugh2008}. Using coronal images during
flares, there are observational reports about detection of the
loop contraction during flares \citep{Liu2009}.

Further, it was shown by \cite{pvk1990} that in force-free
fields a high non-potentiality implies weaker magnetic tension,
which in turn implies a larger vertical extension of the field
due to lower magnetic pressure gradient. Conversely, the
release of free magnetic energy during flare implies a loss of
magnetic non-potentiality leading to a decrease in the vertical
extension of the field or shrinkage \citep{Forbes1996}.

The non-linear force-free field (NLFFF) extrapolations of the
NOAA 10930 active region by \cite{Schrijver2008} show the
non-potentiality of this active region in the form of a twist
flux rope structure. As suggested by \cite{pvk1990} and
\cite{Hugh2000}, such a structure will have larger vertical
extension in pre-flare as compared to the post-flare
configuration. The closer the post-flare field approaches to
the potential field configuration the smaller is the value of
inclination difference $\Delta\gamma$ expected. This may give
an explanation for the decrease of dip-shear $\Delta\gamma$
after the flare.


However, in contrast, the increase in the twist-shear $\Delta
\psi$ after the flare also needs an explanation. The opposite
behaviour of twist-shear and dip-shear in relation to the flare
can be understood in the following way. The twist-shear is
dependent on sub-photospheric / photospheric forces, so the
twist-shear will continue to increase independent of coronal
processes like flare. However, the plasma $\beta$ decreases
rapidly above the photosphere and thus there is no non-magnetic
force or shear that is strong enough to change the inclination
of the field lines. Hence inclination will be more responsive
to coronal processes. This may explain why inclination became
more potential after the flare. Hence, dip-shear could be a
better diagnostic of NP above the photosphere.

 In summary, we studied the evolution of twist-shear and
dip-shear in a flaring $\delta$-sunspot using a sequence of
high-quality vector magnetograms spanning the pre-flare and
post-flare phases and found that : (i) the penumbra located
close to the flaring site has high twist-shear and dip-shear as
compared to other parts of the penumbra, (ii) the twist-shear
increases after the flare which was earlier reported by
\cite{Jing2008} also, (iii) the dip-shear however shows a
decrease after the flare, (iv) the twist-shear and dip-shear
are correlated, i.e. pixels with high twist-shear exhibit high
dip-shear, and this correlation is much tighter after the
flare, and (v) distribution of twist-shear and dip-shear
 and its evolution (in figure 5) clearly shows different
 patterns before and after the flare.

This type of behaviour in the twist-shear and dip-shear
parameters will need to be evaluated further in more flares
before it can be understood physically. We plan to carry out
more extensive study of the dip-shear and twist-shear in
existing {\it Hinode} datasets. However, a high-cadence study
of these shear parameters would be possible only with the
upcoming observations from Helioseismic and Magnetic Imager
(HMI) onboard Solar Dynamics Observatory (SDO)
\citep{Scherrer02}.  The present study is important in the
sense that it points the way to a vector-field follow-up to the
results of \cite{Sudol2005}, which established the
line-of-sight field changes during powerful flares.

 In the context of present study, one should bear in mind that
the vector magnetograms derived from the {\it Hinode} SOT/SP
scans, although polarimetrically very precise are very noisy
geometrically. An unwanted consequence of the geometric noise
could be that the flows, specially on long time scales, would
tend to create an appearance of non-potentiality, even if there
was none. This is an important issue which needs to be
addressed sooner than later, considering the widespread use of
SOT/SP magnetic maps as the ``vector magnetograms". We plan to
carry out a detailed study of this effect using simultaneously
observed Spectro-Polarimeter (SP) scan from {\it Hinode} SOT
and vector-magnetograms from HMI onboard SDO.

\acknowledgments We thank the anonymous referee for his/her
valuable comments and suggestions, specially for pointing out
the geometric noise in  SOT/SP scans and its side effects. We
thank Dr. Ron Moore and Dr. Pascal D$\acute{{\rm e}}$moulin for
reading the manuscript. Hinode is a Japanese mission developed
and launched by ISAS/JAXA, collaborating with NAOJ as a
domestic partner, NASA and STFC (UK) as international partners.
Scientific operation of the Hinode mission is conducted by the
Hinode science team organized at ISAS/JAXA. This team mainly
consists of scientists from institutes in the partner
countries. Support for the post-launch operation is provided by
JAXA and NAOJ (Japan), STFC (U.K.), NASA, ESA, and NSC
(Norway). We also would like to thank Dr. Andreas Lagg for
providing his HELIX code used in this study.


\begin{figure}
\begin{center}
\includegraphics[angle=0,scale=.74]{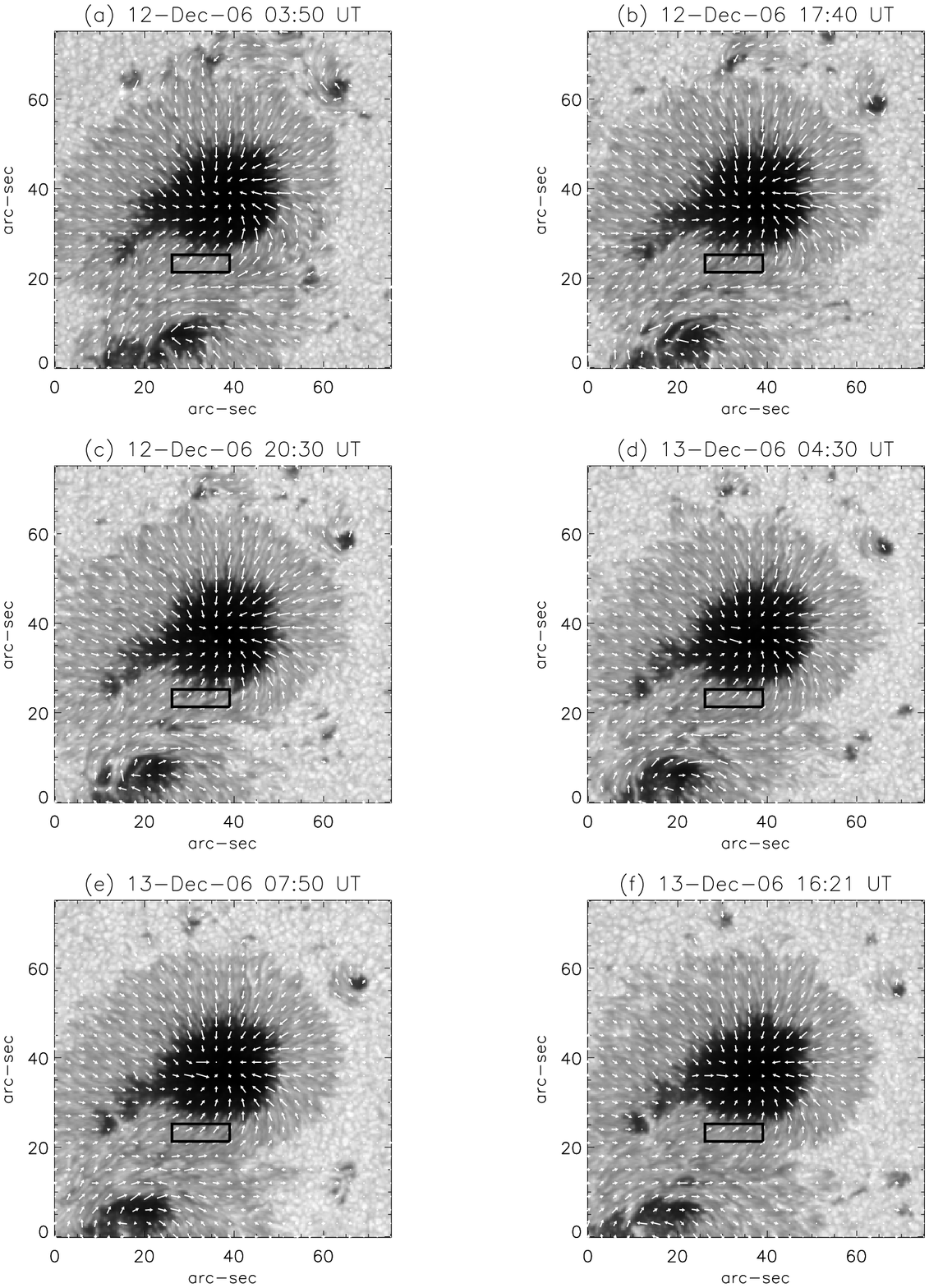}
\end{center}
\caption{ The panels (from top to bottom) show continuum
intensity map of the $\delta$-sunspot in NOAA 10930 during the
times mentioned at the top. The transverse magnetic field
vectors are shown by arrows overlaid upon these maps. The black
rectangle, shown in all
 panels, is the region where we monitor the evolution of twist-shear and dip-shear.} \label{fig:continuum}
\end{figure}

\begin{figure}
\begin{center}
\includegraphics[angle=0,scale=.7]{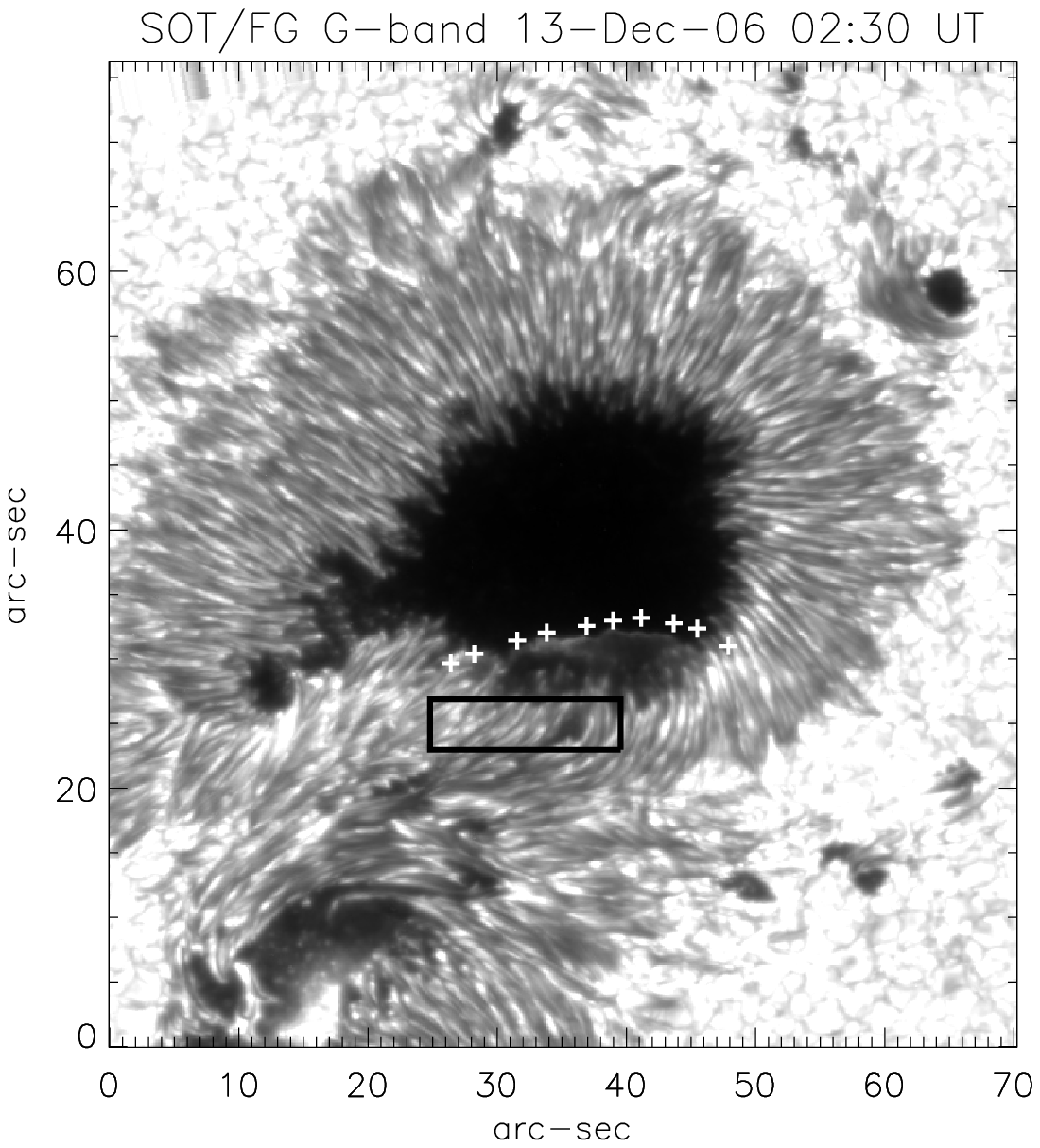}\includegraphics[angle=0,scale=.7]{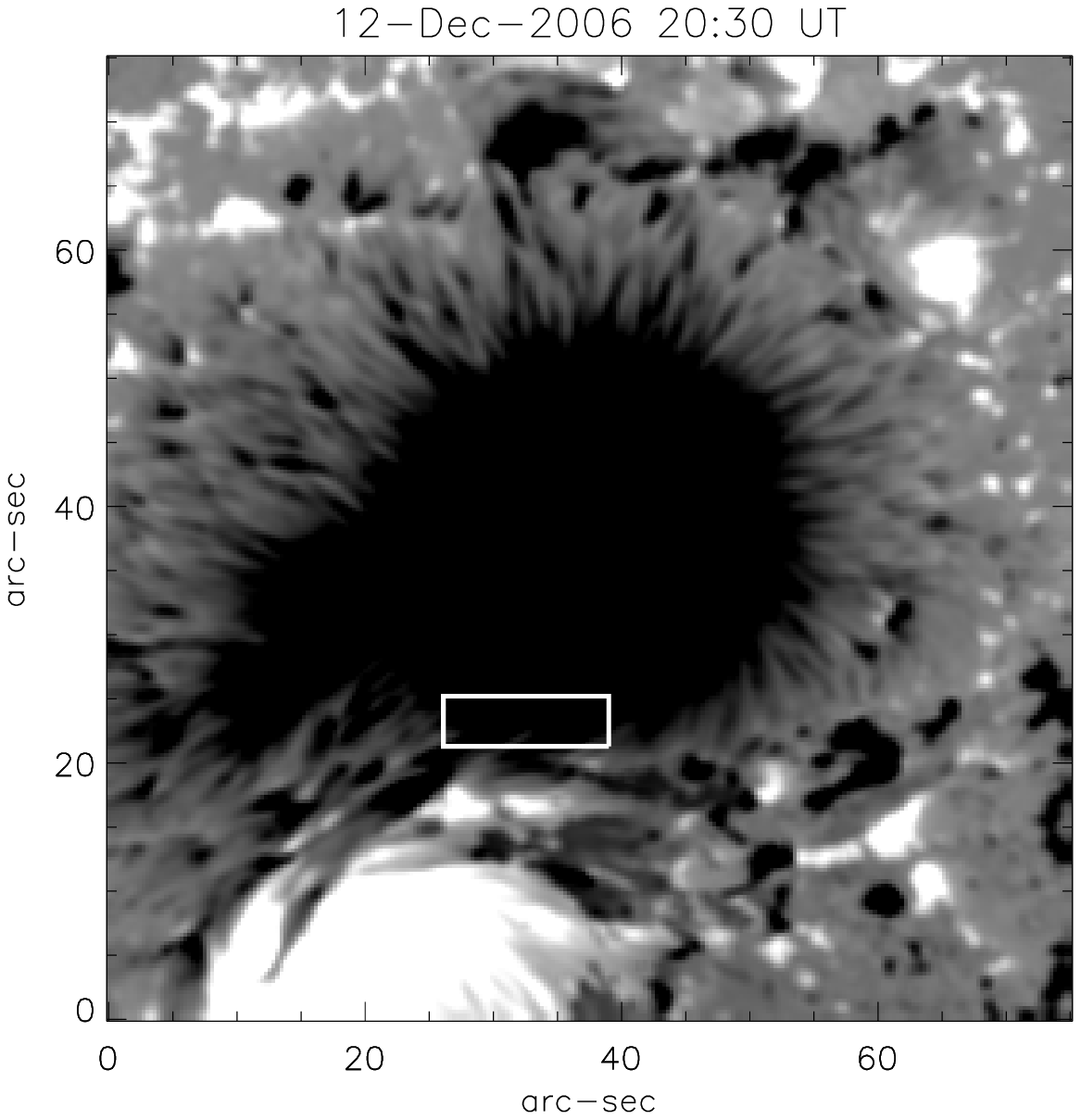}
\end{center}
\caption{ The left panel shows the G-band filtergram of
the $\delta$-sunspot in NOAA AR 10930 during 12 December 2006
02:30 UT, with the location of the flare ribbon marked by `+'
symbols. The flare ribbons sweep across the rectangular box
during 02:20 to 02:26 UT. The right panel shows the map of the
longitudinal field component for this sunspot. The black
(white) rectangle in left (right) panel marks the region where
we monitor the evolution of twist-shear and dip-shear. }
\label{fig:gband}
\end{figure}

\begin{figure}
\begin{center}
\includegraphics[angle=0,scale=.75]{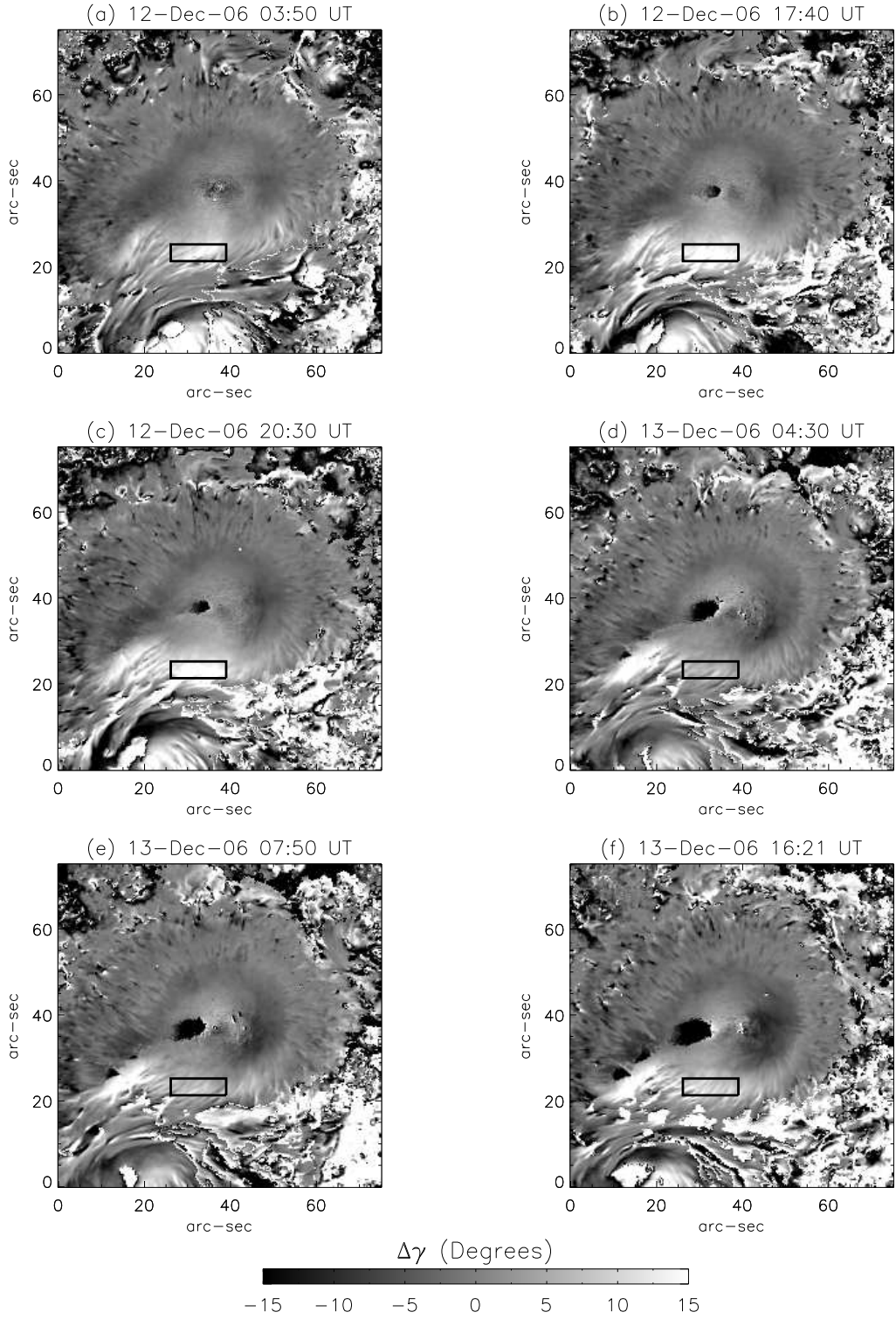}
\end{center}
\caption{The panels (from top to bottom) show the maps of
dip-shear, $\Delta\gamma=\gamma_{o}-\gamma_{p}$ for NOAA AR
10930 at different times. The black rectangle marks the region
where we monitor the evolution of twist-shear and dip-shear.}
\label{fig:dgama}
\end{figure}

\begin{figure}
\begin{center}
\includegraphics[angle=0,scale=.75]{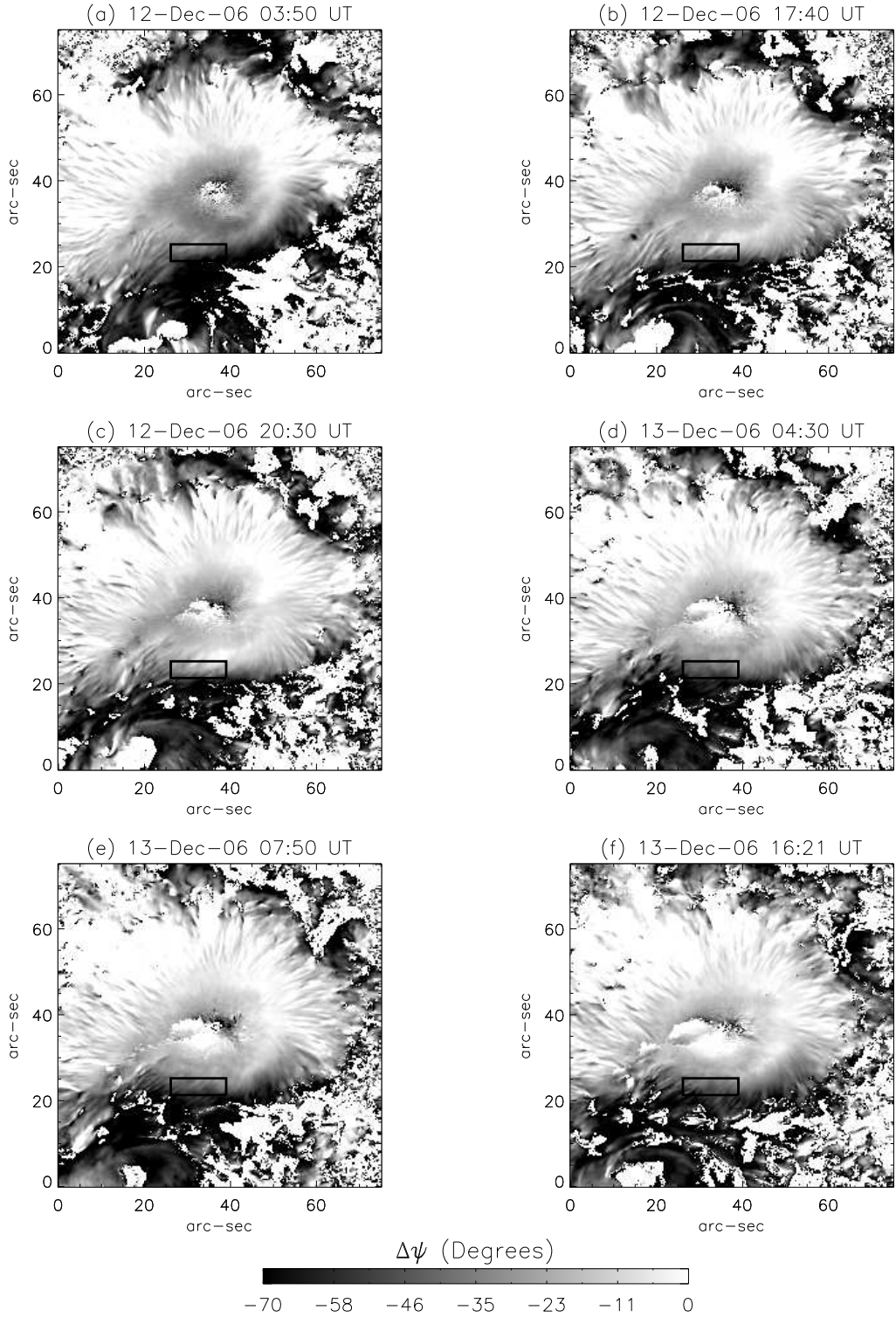}
\end{center}
\caption{The panels (from top to bottom) show the maps of twist-shear, $\Delta\psi=\psi_{o}-\psi_{p}$ for NOAA AR 10930 at different times.
The black rectangle marks the region where we monitor the evolution of twist-shear and dip-shear.}
\label{fig:dgama}
\end{figure}

\begin{figure}
\begin{center}
\includegraphics[angle=0,scale=.72]{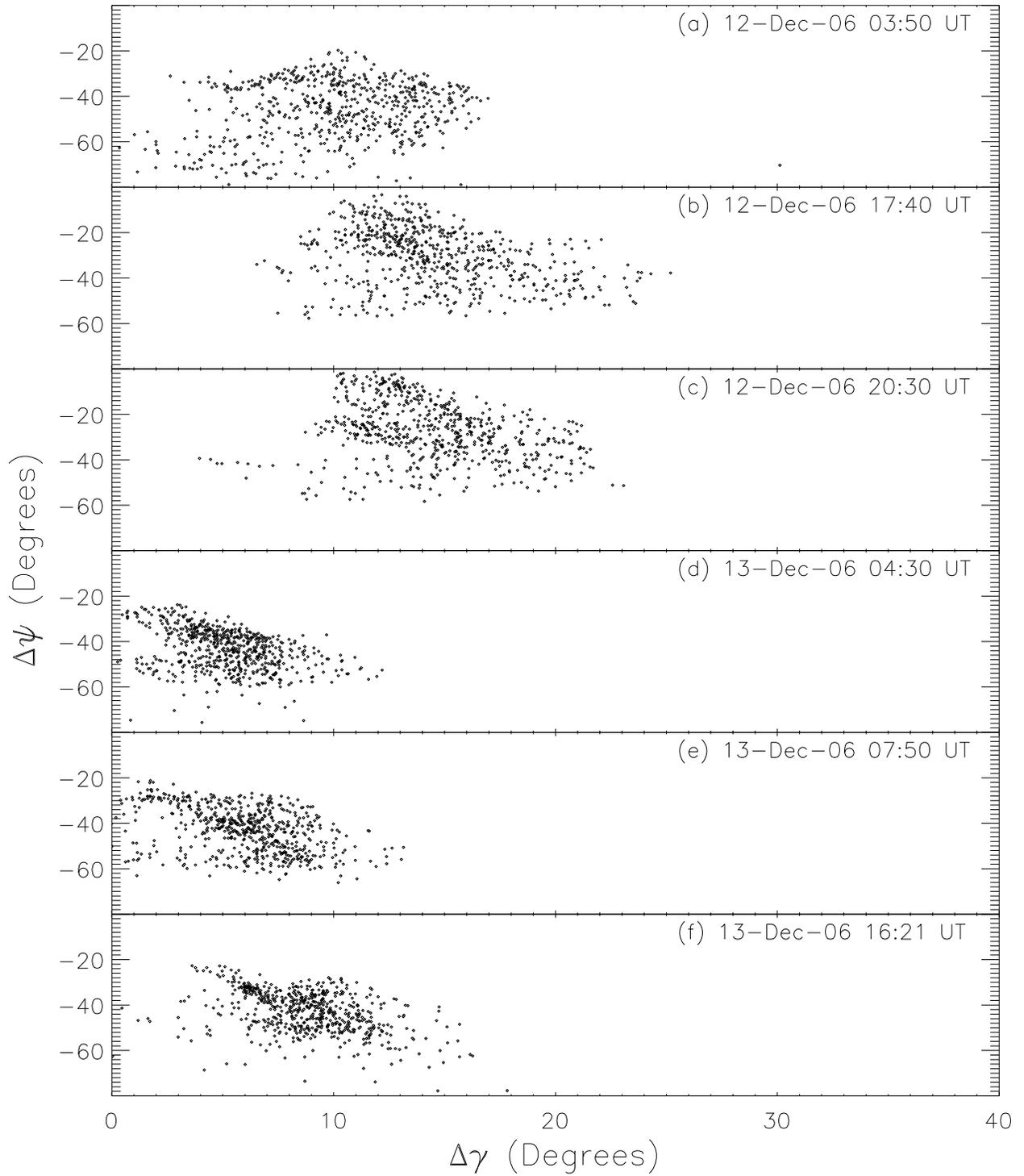}
\end{center}
\caption{ The panels (from top to bottom) show the evolution of
twist-shear $\Delta\psi$ and dip-shear $\Delta\gamma$ inside
the region marked by rectangle in previous figures. }
\label{fig:scatter}
\end{figure}

\end{document}